\documentclass{article}

\usepackage{amsmath,amsfonts,amssymb}
\renewcommand{\Re}{\mathbb R}
 \newcommand{\ii}{\text{\bf\em i}}

\newcommand{\talk}[1]{{\bf #1}} 
\newtheorem{thm}{Theorem}[section]
\newtheorem{conj}{Conjecture}

\begin{document}
\title{Report on GR16, Session A3: \\
Mathematical Studies of the Field Equations}
\author{Lars Andersson\\
Department of Mathematics\\ University of Miami$^\dagger$}
\renewcommand{\thefootnote}{}
\footnotetext{$^\dagger$ Supported in part by the Swedish Natural
Sciences Research Council (SNSRC),  contract no.  R-RA 4873-307 and the NSF,
contract no. DMS-0104402.  Email: larsa\char'100math.miami.edu}
\maketitle
\begin{abstract}
In this report, which is an extended version of that
appearing in the Proceedings of GR16, I will give a summary of the main
topics covered in Session A.3. on mathematical relativity 
at GR16, Durban. 
The summary is mainly based on extended abstracts submitted by the speakers. 
I would like to thank all participants for their contributions and help with
this summary.
\end{abstract}

\section{The Horowitz-Myers Soliton}
According to a conjecture by Maldacena\cite{maldacena:talk},
supported by a growing body of evidence, there is a correspondence between
string theory in Anti--deSitter (AdS) spacetime and a conformal field theory
(CFT) on the boundary of AdS.
Stability considerations motivated by the AdS/CFT conjecture led to the
conjecture that the boundary of asymptotically AdS space--times should be
connected. The Riemannian version of this was proved by Witten and 
Yau\cite{witten:yau} and generalized by Cai and Galloway\cite{cai:galloway}.
The non--supersymmetric version of the
AdS/CFT correspondence, leads to consideration of asymptotically locally AdS
spacetimes with conformal boundary such that one spatial direction is
compactified on a circle.
If we allow nontrivial topology at infinity, the positive mass theorem for
general relativity is not valid in general\cite{horowitz:myers}.
In case of negative cosmological constant,
$\Lambda < 0$, a partial Cauchy surface
in an asymptotically locally AdS $n+1$-dimensional space--time may have any
orientable $n-1$--manifold as boundary at infinity. The Kottler and Nariai
metrics give explicit examples of static
AdS metrics on manifolds with nontrivial topology
at infinity\cite{chrusciel:simon}. These examples all have horizons.

Horowitz and Myers\cite{horowitz:myers}
have found a static solution of the Einstein
equations with negative cosmological constant, $\Lambda < 0$, which is
asymptotically locally AdS and contains no horizon.  I will refer to this
spacetime as the HM soliton.  The HM soliton in $n+1$ dimensions is given by
the metric
\begin{equation}
ds^2 = -r^2dt^2 + \frac{1}{V(r)}dr^2 + V(r)d\phi^2 +r^2
\sum\limits_{i=1}^{n-2}(dy^i)^2 \label{eq:HMmetric}
\end{equation}
where $V(r)= \frac{r^2}{\ell^2} \left ( 1-\frac{r_0^{n}}{r^{n}} \right )$,
$\ell^2=-\frac{n(n-1)}{2\Lambda}$, and $r_0$ is a constant. In order to avoid
an orbifold singularity, $\phi$ must be identified with period
$\beta_0=\frac{4\pi\ell^2}{n r_0}$. After making the $y^i$ coordinates
periodic as well (with arbitrary period), the resulting spacetime has
conformally compact Cauchy surfaces with flat torus conformal
boundary.

Computing the mass of the soliton using a standard method gives the negative
answer $E = - \frac{\beta r_0^2}{4 G_5 \ell}$.
The mass at infinity of metrics asymptotic to the soliton is related via the
AdS/CFT correspondence to the negative Casimir energy of a
non--supersymmetric gauge theory on the boundary. If a non--supersym\-metric
version of the AdS--CFT conjecture holds, then stability considerations imply
that the soliton must be the lowest energy state with the given asymptotic
conditions. This leads to a new positive mass conjecture,
\begin{conj}[Horowitz and Myers]
Spacetimes satisfying a dominant energy condition with $\Lambda < 0$, and
with the asymptotic behavior of the soliton (\ref{eq:HMmetric}), have mass
bounded from below by $E$. The soliton is the unique lowest
mass solution for all spacetimes in its class.
\end{conj}
This conjecture is supported by perturbation calculations up to second 
order\cite{horowitz:myers,constable:myers}.
A further piece of supporting evidence was provided by
the recent work of Galloway,  Surya and Woolgar\cite{galloway:etal:uniqueness},
described by \talk{Greg Galloway} in his talk at session A.3. Galloway et al.
proved that the HM soliton is the
unique static metric in a class of metrics with the given asymptotic
conditions. In order to formulate the main result of Galloway et. al., 
let $(\Sigma, h , N)$ be a solution of the
static vacuum field equations
$
R_{ab} = N^{-1}\nabla_a\nabla_b N +\frac{2\Lambda}{n-1} h_{ab} , 
\Delta N  = -\frac{2\Lambda}{n-1} N ,
$
which is conformal
to a compact manifold with boundary
$(\tilde\Sigma, \tilde h)$ 
with conformal factor $\tilde N = 1/N$ (with $\tilde N = 0$ on
$\partial \tilde \Sigma$.  Then $(\Sigma, h,N)$ is said to satisfy condition
(S) if the level sets $\tilde N = c$ near infinity are weakly convex in the
sense that the principal curvatures are either all positive or all negative.
The mass is computed as an integral over the conformal boundary of the mass
aspect of $(\Sigma, h, N)$, defined as (specializing the Ashtekar--Magnon
conformal mass to the static case)
$\mu =
-\partial^{n-2}
\widetilde R/\partial x^{n-2}|_{x=0}$ (where $x= \tilde N$ and $\tilde R$
is the scalar
curvature function of $(\tilde \Sigma, \tilde h))$.
\begin{thm}
Let $(\Sigma,h,N)$ be a static spacetime as above, satisfying condition (S).
Suppose in addition, that 
(a) The conformal boundary of $(\Sigma,h,N)$ 
is the same as that of the $H\&M$ soliton,
(b) The mass aspect $\mu$ of $(\Sigma,h,N)$ is pointwise negative, and 
(c) The kernel of the inclusion map 
$i_* : \Pi_1(\partial \tilde\Sigma) \to \Pi_1( \tilde\Sigma)$ is generated by
the $S^1$ factor. 
Then the spacetime
$
M^{n+1} = \Re\times \Sigma, \qquad g= -N^2dt^2 \oplus h .
$
determined by $(\Sigma,h,N)$ is isometric to the AdS soliton
(\ref{eq:HMmetric}).
\end{thm}
The proof uses complete null lines constructed using the convexity of the
boundary together with a 
null splitting theorem of Galloway\cite{galloway:nullsplit} to prove that the universal cover of the conformal
compactification $(\tilde\Sigma, \tilde h)$ splits as a metric product
$\Re^{n-2}\times W$ with the metric on $W$ determined uniquely by the field
equations.

\section{Dynamical systems in general relativity}
\talk{Claes Uggla} reviewed some applications of dynamical systems methods
in general relativity. Dynamical
systems formulations arise naturally in the study of Einstein's
field equations for special classes of spacetimes, such as
hypersurface--homogenous and hypersurface-self-similar space\-ti\-mes\cite{uggla:homog}.  Further examples are provided by the geodesic equations
of certain classes of spacetimes, where dynamical systems techniques can be
used to e.g., calculate the null geodesics and the corresponding cosmic
microwave background radiation pattern in a cosmological model\cite{uggla:geodesic2,nilsson:uggla:etal:isotropic}, and the
density perturbations of
spatially homogenous models.
The classical example of application of
dynamical systems methods to the Einstein equations is the study of spatially
homogenous (Bianchi) space--times, with or without matter.  The Bianchi
family of spacetimes contains the Friedman spacetimes as a special case, and
gives a rich set of examples of the influence of matter and geometry on the
early and late time evolution of space--times. It is a natural problem to
analyze completely the behavior of the resulting family of dynamical systems.

The book on dynamical systems in cosmology 
edited by Ellis and Wainwright \cite{ellis:wainwright:book} gives a
good overview of the state of the art of this approach, some years ago.  As
an outgrowth of that work, substantial progress has been made towards
understanding the Bianchi systems both near the initial singularity and in
the asymptotically expanding direction.  
\talk{Hans Ringstr\"om}\cite{ringstrom:blowup,ringstrom:bianchiIX}
proved that Bianchi VIII and IX have oscillating singularity, which is an
important step in understanding the full BKL picture for these models. An
important step in the analysis was to prove that Bianchi II is an attractor
for Bianchi IX. This sort of hierarchical behavior in the Bianchi models,
is an important organizing principle also for further work on
systems with lower dimensional isometry groups. 
Ringstr\"om has also, as he described in his  talk, been able to analyze in
detail\cite{ringstrom:future}
the asymptotic behavior of the expanding vacuum Bianchi class A
spacetimes
(this excludes Bianchi IX). For the most complicated of these,
Bianchi VIII, the result is that the geometry is asymptotically Taub, with
precise control on the asymptotic behavior of the commutator variables. For
Bianchi VII$_0$, the system approaches (in terms of scale free variables), a
flat VII$_0$ spacetime.
The Cauchy surfaces of  these spacetimes are all Seifert fibered.  The
results show that the scale free geometry of the spatial sections in these
spacetimes collapse along the fiber direction. The result is consistent with
the picture for the limiting behavior in the asymptotic expanding direction
developed by M. T. Anderson\cite{anderson:asympt}
Analyzing the 
hierarchical structure of systems with small symmetry group, such as 
Gowdy models,
or without
symmetry, 
is likely to be a crucial step in understanding the global
dynamics of the Einstein equations.

\section{Well-posed forms of the 3+1 trace--de\-coupled Einstein equations}
The standard ADM form of the Einstein evolution
equations fails to be well posed both from the mathematical and numerical
points of view. Due to the demand for stable numerical procedures
for solving the Einstein evolution equations, a great deal of interest has
recently been focussed on well posed hyperbolic formulations of the Einstein
equations. This is of particular importance in solving boundary value
problems. 
\talk{Oscar Reula} discussed various well--posed reformulations of
the Einstein equations. It is possible to derive a well--posed form of the
standard 3+1 Einstein equations\cite{york79} by reducing to first order
form, densitizing the lapse function and adding combinations of the
constraints to the evolution equations\cite{frittelliletter,frittelli-reula99}.
An alternative approach, which recently has been shown to possess striking
computational advantages over the standard form\cite{BS99}, is to decompose
the fundamental variables in order to extract the trace of the extrinsic
curvature and the determinant of the 3-metric with the purpose of evolving
them independently from the remaining components.  This trace--decoupled form
of the Einstein equation turns out, oddly enough, to be ill--posed. Reula and
Frittelli have used techniques similar to those used for the standard
evolution equations, to obtain versions of the 3+1 Einstein equations which
are both trace-decoupled and well posed.
This well posed version propagates the constraints in a stable manner, which
is relevant to unconstrained evolution. It uses essentially the same
variables as the original system, but requires a gauge condition, namely that
the lapse to be proportional to the determinant of the intrinsic geometry of
the surfaces.  

\section{Critical phenomena}
Critical phenomena in general relativity have been extensively studied since
the discovery by Choptuik of universal behavior in the collapse of a
self--gravitating scalar field. Since then a number of matter models, such as
charged self--gravitating scalar fields, Yang--Mills and dilaton fields have
been studied and a systematic understanding of these phenomena has been
gained, in particular through the study of the stability of models with
continuous or discrete self--similarity.  We now have a good
qualitative understanding of criticality in 
gravitational
collapse: there is an intermediate (codimension-1) attractor of the evolution
flow in Phase Space, sitting on the threshold of black hole formation. The
symmetries of that attractive solution determine the type of criticality:
`type I' if the solution is static, `type II' when it is self-similar.

The Einstein--Vlasov system describes the evolution of a statistical
ensemble of non-interacting particles coupled to gravity through their
average properties.
\talk{Jose M. Martin--Garcia}
described joint work with Carsten Gundlach on critical phenomena in the
Einstein--Vlasov system, which is
important in Critical
Phenomena Theory because it is the only one where type II critical phenomena
have not yet been found.

There are two different ways of checking for the existence of criticality. On
the one hand, it is possible to evolve numerically families of initial
conditions and study the collapse threshold, looking for intermediate
attractors and universality. On the other hand, one can construct static or
self-similar solutions and then investigate whether they are codimension-1
linearly stable.

There are two relevant works in this problem, both using the method of
evolution of initial conditions: Rendall, Rein and Schaeffer\cite{RRS98}
have not found criticality and suggest that, if there is any, it must be type
I. Olabarrieta and Choptuik\cite{OlCh01} confirm the apparent nonexistence
of type II phenomena and have found some signs of type I criticality, though
not conclusive because there is not universality in the critical exponents.
Gundlach and Martin--Garcia look for type II criticality in the massless
Einstein-Vlasov system using the second method. 

In his talk Martin--Garcia
reported on the first step: the numerical construction of self-similar
solutions, which is an interesting problem in itself.
The situation resembles what is already known analytically for static
solutions of the same system\cite{Rein93}.  There is a free function of two
variables, representing different distributions of the energy-momentum among
the particles of the system, but all giving rise to the same gravitational
field. Therefore a candidate critical solution can not be isolated. It is
likely that a similar result in the massive case could explain the observed
non-universality.

Recent numerical investigations\cite{CP,HO} of near-critical scalar field
collapse in 2+1 dimensional AdS spacetime show a continuous self-similar
(CSS) behaviour near the central singularity and power law scaling for the
black hole mass. Garfinkle\cite{gar} found that this behaviour is well
approximated, at intermediate times, by a member of a one-parameter family of
exact CSS solutions to the equations with vanishing cosmological constant
$\Lambda$. However the extension of these solutions to $\Lambda\neq 0$,
necessary to explain black hole formation (as vacuum black holes exist only
for $\Lambda < 0$) is a hard problem which has not been solved up
to now.

\talk{G\'erard Cl\'ement} in his talk discussed joint work with Alessandro
Fabbri\cite{clement:quasi} 
on quasi-CSS 2+1 dimensional scalar field spacetimes.
They have derived by a limiting process a new class of $\Lambda = 0$ CSS
solutions
\begin{eqnarray}
\label{new2} ds^2 & = & dudv - (-u)^{2/(1+c^2)} d\theta^2,
\nonumber \\ \phi & = & -\frac{c}{1+c^2}\ln(-u)
\end{eqnarray}
These 
present for $c^2 \le 1$ a point singularity ($u=0$, $v=+\infty$), and for
$c^2 < 1$ a null line singularity $u = 0$.  They can be extended to exact
solutions of the $\Lambda = -l^{-2} < 0$ equations by means of the ansatz
\begin{equation}
\label{anext}
ds^2 = e^{2\sigma(x)}dudv - (-u)^{\frac2{c^2+1}}\rho^2(x)d\theta^2, \quad
\phi = -\frac{c}{c^2+1}\ln|u| + \psi(x)\,,
\end{equation}
where $x = uv$, with the initial conditions $\rho(0) = 1, \, \sigma(0) =
\psi(0) = 0$. These extended solutions inherit the CSS behaviour close to the
central singularity, and have the correct AdS behaviour at spatial infinity.

To show that these quasi-CSS solutions are indeed threshold solutions for
black hole formation, one studies their linear perturbations.
The case $c=0$ is at the same time simple and interesting. In this case the
scalar field decouples, $\phi=0$, and the corresponding quasi-CSS solution is
the vacuum BTZ\cite{BTZ} solution. The linear perturbation equations with
appropriate
boundary conditions lead to a unique solution, with eigenvalue
$k=2$. This solution is found to be an exact linearization in $M$ of the BTZ
black hole.
In the case of genuine scalar perturbations, one finds
two possible modes $k_a = c^2 + 3/2$ and $k_b = c^2 + 2$. For the $a$ mode
the singularity and the apparent horizon appear simultaneously on the null
line $u = 0$ at the time $v=0$ and evolve in the region $v>0$ in a physically
meaningful way, while for the $b$ mode the singularity still appears for $v =
0$, but the apparent horizon seems to be eternal, as in the case of the
static BTZ black hole. The $b$ mode, which cannot
describe actual gravitational collapse with regular initial conditions,
appears to be unphysical.
The critical exponent is related to the eigenvalue $k$ by $\gamma=1/k$. In
the BTZ case $c=0$  Cl\'ement and Fabbri  obtain $\gamma=1/2$.
In the scalar field case, choosing
the critical value $c^2 = 1$ they obtain for the $a$ mode $\gamma = 0.4$. This
value differs signicantly from those found in the numerical
analysis\cite{CP,HO}, i.e. $\gamma \sim 1.2$ and $\gamma\sim 0.81$,
showing that further work is needed in order to clarify this issue.

\section{Isolated systems}
It is an open problem to construct a nonflat vacuum spacetime,
which has a regular conformal infinity in the sense of Penrose. The conformal
compactification of a nonflat spacetime must have a singularity at spatial
infinity $\ii_0$, due to the slow falloff of the Weyl tensor at $\ii_0$.
\talk{Sergio Dain} described joint work with Helmut Friedrich\cite{Dain99}, 
which constructs a large class of asymptotically flat initial data with
non-vanishing mass and angular momentum for which the metric and the
extrinsic curvature have asymptotic expansions at space-like infinity in
terms of powers of a radial coordinate. These asymptotic expansions are of
the form
$
\tilde h_{ij}\sim (1+\frac{2m}{\tilde r})\delta_{ij}+\sum_{k \geq2}
\frac{\tilde h^k_{ij}}{\tilde r^k}, 
\tilde \Psi_{ij}\sim \sum_{k \geq2} \frac{\tilde \Psi^k_{ij}}{\tilde r^k},
$
where $\tilde h^k_{ij}$ and $\tilde \Psi^k_{ij}$ are smooth function on the
unit 2-sphere (thought as being pulled back to the spheres $\tilde{r} =
const.$ under the map $\tilde{x}^j \rightarrow \tilde{x}^j/\tilde{r}$).  In
earlier work\cite{Beig80,Dain01b}, stationary data were
shown to admit expansions of this type.

In the study of
isolated system, an important and difficult problem is to give an unambiguous
definition of angular momentum. 
The difficulty lies in the fact that without introducing extra structure,
there is no unambiguously defined Poincar\'e subgroup of the asymptotic
isometries of an isolated system. This is known as the supertranslation
ambiguity.
This ambiguity is known as supertranslation ambiguity.  In order to
provide with an unambiguous notion of angular momentum in radiative
spacetimes, it is essential to make use of a reference frame system, which
embodies the notion of rest frame\cite{Winicour80}.
\talk{Osvaldo Moreschi} presented a definition of angular momentum
for radiative spacetimes which does not suffer from supertranslation
ambiguity.  The definition uses the construction of so called `nice
sections' at $\mathcal I^+$. The defining equations for these
have recently been proved\cite{Moreschi98,Dain00} to have solutions
in terms of a 4-parameter family of translations.  The solutions have the
expected physical properties of a rest frame.
The nice section construction singles out a Poincar\'e structure from the
infinite dimensional BMS group. In particular, given a fixed observational
point $p$ at $\mathcal I^+$, there is precisely a 3-degree of freedom of
spacelike translations which generate all the nice sections that contain $p$.
In contrast, without this constructions there is an infinite dimensional
family of general sections that contain $p$, one for each supertranslation.
The condition that the spatial part of the Bondi 4--momentum is zero singles
out a one parameter family of nonintersecting\cite{Dain00} sections $S_{\tt
cm}$ of $\mathcal I^+$, which define a center of mass frame and can be used to
describe the detailed asymptotic structure of the spacetime.  The angular
momentum is defined in terms of a charge integral over the $S_{\tt cm}$.

Detectors of gravitational waves can be considered as observers at future
null infinity. The construction of center of mass sections, provides for each
point $p$ at scri a prescription that singles out the unique center of mass
section containing $p$, together with the appropriate rest frame to calculate
intrinsic quantities like angular momentum and multipole moments, which are
important for the description of the gravitational waves that one wishes to
detect.

\talk{John L. Friedman}
described described in his talk joint work with K\=oji Ury\=u and Masaru
Shibata. They consider compact binary systems, modeled as vacuum or
perfect-fluid spacetimes with a helical Killing vector $k^a$, heuristically,
the generator of time-translations in a corotating frame.  Systems that are
stationary in this sense are not asymptotically flat, but have asymptotic
behavior 
corresponding to equal constant fluxes of ingoing and outgoing
radiation.  For black-hole binaries, a rigidity theorem implies that the
Killing vector lies along the horizon's generators, and from this one can
deduce the zeroth law (constant surface gravity of the horizon).
Remarkably, although the
mass and angular momentum of such a system are not defined, 
a finite Noether charge is defined on any sphere enclosing the matter 
and black holes, its value the same on each such sphere.  An exact first law
relates the change in this charge to the changes in the vorticity, 
baryon mass and entropy of the fluid and in the area of black holes.
Modeling a binary system by a spacetime with a helical Killing vector
is accurate if the energy emitted in a dynamical time is small compared 
to the kinetic energy of the system. 
%

Other approaches use a post--Newtonian
approximation, or conformally flat spacelike slices that satisfy a truncated
set of field equations. Like the post-Newtonian spacetimes, these conformally
flat spacetimes are nonradiative and asymptotically flat.
In the isolated horizon framework, for a horizon with a single Killing
vector, one shows the existence of a charge $E$ defined on an isolated
horizon for which $\delta E = \kappa \delta A$. The first
law for black holes with helical symmetry, in contrast, relates this change
in the black-hole charges to the changes in the Noether charge of a sphere
surrounding all black holes and all matter and to the changes in the entropy,
baryon number, and circulation of the fluid.  The existence of such a first
law depends precisely on what is {\it not} assumed in the isolated horizon
framework: a globally defined Killing vector.

\subsection{Isolated horizons} A closely related topic, isolated horizons
(IH) was discussed by \talk{G. Date}. Here again
the work is motivated by the question if it is possible to have
analogues of the usual laws of BH mechanics in spacetimes more
general than the usual stationary asymptotically flat black holes.

G. Date stated the following
results\cite{date:killing,date:isolated}: 
A Killing horizon (KH) is always an IH, and
an IH is a KH provided the IH admits a neighbourhood which is
{\it{either}} stationary {\it{or}} has two commuting Killing
vectors. Further, a vacuum, non-rotating IH  necessarily admits an EH and
for a rotating IH , an EH exists if $\int {\mathcal{E}} + \pi
\bar{\pi} -2 \pi < 0$. Otherwise, the IH is accessible from infinity. An
EH could exist but the IH will be `outside' of it.

\subsection{Stars}
The Einstein equations for stationary axisymmetric perfect fluid spacetimes
reduces to an elliptic system, known as the Ernst equations. These
spacetimes are commonly studied as models of isolated stars in general
relativity.

\talk{Christian Klein} presented joint work with J\"org Frauendiener on exact
solutions of the stationary axisymmetric Einstein equations for perfect
fluid matter. A boundary value problem for the Ernst equations is solved
using Riemann--Hilbert techniques.
As an application he presented
the solution for a family of counter-rotating dust
disks\cite{klein:prl2,klein:prd4} which contains the rigidly rotating dust 
disk\cite{neugebauermeinel1} as a limiting case.

\talk{H. Pfister} considered existence and non--existence results proved using
Banach space fixed point techniques\cite{pfister1}.

\section{Variational formulation}
\talk{Naresh Dadhich} and \talk{Niall \'O Murchadha} gave talks
relating to the variational formulation of general relativity.
Dadhich considered  the conditions under which one can deduce the field
equations from the equations of motion for particles.
In the context of electomagnetic field, this question was, according to
Dyson \cite{dyson}, first
considered by Feynman. By considering commutation
relations between position and velocity, he could obtain the Bianchi set
of the equations. This gave rise to considerable activity in this
direction involving rederivations, generalizations and extensions. However
the attention centred on the Bianchi set.
Singh and Dadhich\cite{singh:dadhich2} derived the full set of the Maxwell equations
by demanding the law of motion to be linear in velocity and
derivable from a Lagrangian. A similar procedure has been
applied to the Einstein equations\cite{singh:dadhich}. 
In this case, the fact that the
particle law in general relativity is quadratic in the 4--velocity leads to
the Einstein equations. This 
is a novel way of deducing the field equations from the particle law
of motion.

\talk{Niall \'O Murchadha} 
described joint work with Julian Barbour and Brendan Foster\cite{barbour:murch}, where a
derivation of General Relativity was given which does not depend on the
relativity principle. The fundemental idea is to choose superspace (the
space of all riemannian three geometries) as the configuration space and to
consider a parametrised curve in this configuration space, with parameter
$\lambda$.
\'O Murchadha et al. consider a generalized
Jacobi-type action of the form $\int
d\lambda\int dx^3 \sqrt{gPT}$ on such curves where one first integrates on
each given three manifold and then integrates along the curve. `$P$' is the
potential which is a scalar on the three manifold but otherwise arbitrary,
while
`$T$' is the generalized kinetic energy which means that it is a quadratic
function of $\partial g_{ij}/ \partial \lambda$.
It turns out that in order for the constraints of the theory to be
propagated, it is necessary to choose the potential energy term to be the
three scalar curvature and the kinetic energy to be defined with the DeWitt supermetric.
This recovers the Baierlein, Sharp, Wheeler
parametrised action for G.R. as the unique self-consistent Jacobi-type
action on
superspace.

\section{Singularities}
There were several talks concerned with the structure of cosmological and
black hole singularities.

An important development has been the application
of Fuchsian techniques to construct families of asymptotically velocity term
dominated (AVTD) cosmological singularities\cite{rendall:kichenassamy,andersson:rendall:quiescent}.
\talk{M. Narita} discussed his work \cite{narita:etal:gowdy},
applying the Fuchsian method to Gowdy
spacetimes with stringy matter, showing that these have AVTD
singularities. Stringy gravity has receieved a great deal of attention
recently due to the work of Damour and Henneaux\cite{damour:henneaux:homog},
who showed that for stringy gravity in
of no symmetries in spacetime dimension 11 and higher,
one expects mixmaster type
behavior.

\talk{Brien Nolan} has studied the gravitational collapse of
spherically symmetric thick shells admitting a homothetic Killing
vector field under the assumption that the energy momentum tensor
corresponds to the absence of a pure outgoing component\cite{nolan}. 
The stability of the Cauchy horizon against
linear perturbations is investigated. The behaviour of a massless scalar
field $\phi$ propagating on a background space-time is studied. The
following boundary conditions are imposed: (i) finiteness of the flux of
$\phi$ as measured by any time-like observer crossing $N$ and (ii)
finiteness of the flux of $\phi$ in the ingoing null direction measured
at ${\mathcal J}^-$. Imposing these boundary conditions, it is found that
the flux of $\phi$ as measured by any time-like (geodesic) observer
crossing the Cauchy horizon is always finite. This is in contrast to the
case of Cauchy horizons inside black holes. Those arising here always
lie outside any possible event horizon.
This indicates that the Cauchy horizon may be stable against linear and
possibly non-linear gravitational perturbations, and so members of this
class of space-times may provide physically realisable examples of
naked singularities.

\talk{A. Beesham} also discussed naked singularities. In joint work with
S.G.~Ghosh and 	R.V.~Saraykar\cite{beesham} he has studied
gravitational collapse of radiation shells  in a non self-similar
higher dimensional spherically symmetric spacetime.
Strong curvature naked
singularities were found to form for a highly inhomogeneous collapse,
violating the cosmic censorship conjecture for this class of models.

\talk{Deborah Konkowski} reviewed the status of two instability conjectures for
Cauchy horizons\cite{konkowski}.

\section{Miscellaneous}
\subsection{Lanczos and Bel--Robinson}
In his talk, \talk{Brian Edgar}
reviewed his work with F. Andersson on the Lanczos
potential for the Weyl tensor\cite{andersson:edgar1,andersson:edgar2}.
The existence proof for the Lanczos potential
is now on firm foundation, and the construction of the potential in several
situations with symmetry is understood. Unfortunately, a really interesting
application of this intriguing object is still lacking.

\talk{Jose M. Senovilla} discussed his work on Bel--Robinson squares\cite{senovilla}. This work
generalizes and unifies the earlier literature on Bel--Robinson and Bel
tensors. The Bel--Robinsons tensor was introduced as a candidate for a
``stress--energy'' tensor for the gravitational field, and in 3+1
dimensional vacuum spacetimes, this
symmetric, trace--free 4--tensor is divergence free and positive definite in
a certain sense. The Bel tensor plays a
similar role for Einstein equations with matter. The work of Senovilla gives
a systematic procedure for constructing similar objects in higher
dimensional spacetimes.
\subsection{Distributional Geometry}
\talk{James Vickers} and \talk{Michael Kunzinger} both in their talks
discussed distributional geometry and
its relation to the cosmic censorship hypothesis. Kunzinger concetrated on
applications of ideas from Colombeau theory in distributional
geometry, while Vickers concentrated on some examples related to cosmic
censorship.

The study of singular spacetimes by distributional methods faces the
fundamental obstacle of the inherent nonlinearity of the field equations.
Staying strictly within the distributional (in particular: linear) regime,
excludes a number of physically interesting
examples such as strings and point particles\cite{gt}.

In recent years, several authors have
therefore employed nonlinear theories of generalized functions, in
particular Colombeau's theory of generalized functions in order to derive a
suitable mathematical framework for a general ``nonlinear distributional
geometry'' adapted to the needs of general relativity\cite{cvw,penrose,geo2,vickersESI}.

Under the influence of these applications in general relativity the
nonlinear theory of generalized functions itself has undergone a rapid
development lately, resulting in a diffeomorphism invariant global theory of
nonlinear generalized functions on manifolds\cite{found,vim,ndg}.  
In particular, a generalized pseudo-Riemannian geometry
allowing for a rigorous treatment of generalized (distributional) spacetime
metrics has been developed.

According to the Cosmic Censorship hypothesis realistic singularities
should be hidden by an event horizon. However there are many examples
of physically realistic spacetimes which are geodesically incomplete,
and hence singular according to the usual definition, which are not
inside an event horizon.

Many of these counterexamples to the cosmic censorship conjecture have
a curvature tensor which is reasonably behaved (for example bounded or
integrable) as one approaches the singularity. James Vickers gave examples of
a class of weak
singularities which may be described as having distributional
curvature\cite{cvw}.

The propagation of test fields on spacetimes with
weak singularities has also been investigated.
A class of singularities\cite{vw1,vw2} which
do not disrupt the Cauchy development of test fields and result in
spacetimes which satisfy Clarke's criterion of `generalised
hyperbolicity' has been found.
Vickers argued that points which are well behaved in
this way, and where Einstein's equations make sense distributionally,
should be regarded as interior points of the spacetime rather than
counterexamples to cosmic censorship.

\subsection{The a--boundary}
\talk{Susan M. Scott} discussed the application of the a--boundary
construction to
cosmic censorship.

The singularity theorems of Hawking and Penrose prove
the existence of incomplete causal geodesics in space-times which satisfy
quite general physical conditions.
It has long been conjectured that
these incomplete causal geodesics would terminate at curvature
singularities, but this has never been proven.
recently established a theorem which goes a considerable distance towards
closing this gap left by the singularity theorems.
By application of the abstract boundary (a-boundary) construction of
Scott and Szekeres\cite{scott:szekeres},
to strongly causal space-times, Ashley and Scott have
shown that strongly causal space-times satisfying the conditions of the
singularity theorems will always have an a-boundary essential singularity.

\subsection{The Ehlers group}
\talk{Marc Mars} discussed his work on the Ehlers group\cite{mars}. The
Ehlers group, which is a symmetry of
the Einstein vacuum field equations for strictly stationary spacetimes,
is considered by Mars from a spacetime
perspective. In this setting, the Ehlers group becomes a subgroup of the
infinite dimensional group
of transformations that maps Lorentz metrics into Lorentz metrics.
Mars has analyzed the global conditions which are required on the spacetime for
the existence of the Ehlers group and has found the  transformation law for
the Weyl tensor under Ehlers transformations.
This allows one to study  where, and under
which circumstances, curvature
singularities in the transformed spacetime will arise. As an application, one
finds a local characterization of the Kerr-NUT metric.

\subsection{Composite spacetimes}
\talk{Katsuhito Yasuno} considered the
dynamics of spatially compact `composite' spacetimes\cite{yasuno}.
Motivated by Thurston's geometrization conjecture, it is interesting to look
for examples of spacetimes with more complicated topology than is allowed by
the spatially homogenous models.
The composite spacetimes are
built from the spatially compact locally homogeneous vacuum spacetimes which
have two commuting
local Killing vector fields and are homeomorphic to torus bundles over the
circle.


\begin{thebibliography}{10}

\bibitem{maldacena:talk}
J.~Maldacena.
\newblock Plenary talk at gr16.
\newblock In {\em GR16 Proceedings}, 2001.

\bibitem{witten:yau}
Edward Witten and S.-T. Yau.
\newblock Connectedness of the boundary in the {A}d{S}/{C}{F}{T}
  correspondence.
\newblock {\em Adv. Theor. Math. Phys.}, 3(6):1635--1655 (2000), 1999.

\bibitem{cai:galloway}
Mingliang Cai and Gregory~J. Galloway.
\newblock Boundaries of zero scalar curvature in the {A}d{S}/{C}{F}{T}
  correspondence.
\newblock {\em Adv. Theor. Math. Phys.}, 3(6):1769--1783 (2000), 1999.

\bibitem{horowitz:myers}
Gary~T. Horowitz and Robert~C. Myers.
\newblock Ad{S}-{C}{F}{T} correspondence and a new positive energy conjecture
  for general relativity.
\newblock {\em Phys. Rev. D (3)}, 59(2):026005, 12, 1999.

\bibitem{chrusciel:simon}
Piotr~T. Chru{\'s}ciel and Walter Simon.
\newblock Towards the classification of static vacuum spacetimes with negative
  cosmological constant.
\newblock {\em J. Math. Phys.}, 42(4):1779--1817, 2001.

\bibitem{constable:myers}
Neil~R. Constable and Robert~C. Myers.
\newblock Spin-two glueballs, positive energy theorems and the
  {A}d{S}/{C}{F}{T} correspondence.
\newblock {\em J. High Energy Phys.}, (10):Paper 37, 30 pp. (electronic), 1999.

\bibitem{galloway:etal:uniqueness}
G.J. Galloway, S.~Surya, and E.~Woolgar.
\newblock A uniqueness theorem for the ads soliton, 2001.
\newblock hep-th/0108170.

\bibitem{galloway:nullsplit}
Gregory~J. Galloway.
\newblock Maximum principles for null hypersurfaces and null splitting
  theorems.
\newblock {\em Ann. Henri Poincar\'e}, 1(3):543--567, 2000.

\bibitem{uggla:homog}
Ulf~S. Nilsson and Claes Uggla.
\newblock Hypersurface homogeneous and hypersurface self-similar models.
\newblock {\em Classical Quantum Gravity}, 14(7):1965--1980, 1997.

\bibitem{uggla:geodesic2}
Ulf~S. Nilsson, Claes Uggla, and John Wainwright.
\newblock A dynamical systems approach to geodesics in {B}ianchi cosmologies.
\newblock {\em Gen. Relativity Gravitation}, 32(10):1981--2005, 2000.

\bibitem{nilsson:uggla:etal:isotropic}
U~S. Nilsson, Uggla C., J.~Wainwright, and W.~C. Lim.
\newblock An almost isotropic microwave temperature does not imply an almost
  isotropic universe.
\newblock {\em The Astrophysical Journal Let.}, page 521, 1999.

\bibitem{ellis:wainwright:book}
J.~Wainwright and G.~F.~R. Ellis, editors.
\newblock {\em Dynamical systems in cosmology}.
\newblock Cambridge University Press, Cambridge, 1997.
\newblock Papers from the workshop held in Cape Town, June 27--July 2, 1994.

\bibitem{ringstrom:blowup}
H.~Ringstr{\"o}m.
\newblock Curvature blow up in {B}ianchi {V}{I}{I}{I} and {I}{X} vacuum
  spacetimes.
\newblock {\em Classical Quantum Gravity}, 17(4):713--731, 2000.

\bibitem{ringstrom:bianchiIX}
Hans Ringstr{\"o}m.
\newblock The {B}ianchi {I}{X} attractor.
\newblock {\em Ann. Henri Poincar\'e}, 2(3):405--500, 2001.

\bibitem{ringstrom:future}
H.~Ringstr{\"o}m.
\newblock The future asymptotics of {B}ianchi {VIII} vacuum solutions.
\newblock {\em Classical Quantum Gravity}, 18(18):3791--3824, 2001.

\bibitem{anderson:asympt}
M.~T. Anderson.
\newblock On long-time evolution in general relativity and geometrization of
  3-manifolds.
\newblock {\em Comm. Math. Phys.}, 222:533--567, 2001.

\bibitem{york79}
Jr. James W.~York.
\newblock Kinematics and dynamics of general relativity.
\newblock In Larry~L. Smarr, editor, {\em Sources of gravitational radiation},
  pages 83--126, Cambridge-New York, 1979. Cambridge University Press.

\bibitem{frittelliletter}
Simonetta Frittelli and Oscar Reula.
\newblock First-order symmetric hyperbolic einstein equations with arbitrary
  fixed gauge.
\newblock {\em Phys. Rev. Lett.}, 76:4667--4670, 1996.

\bibitem{frittelli-reula99}
Simonetta Frittelli and Oscar~A. Reula.
\newblock Well-posed forms of the $3+1$ conformally-decomposed {E}instein
  equations.
\newblock {\em J. Math. Phys.}, 40(10):5143--5156, 1999.

\bibitem{BS99}
Thomas~W. Baumgarte and Stuart~L. Shapiro.
\newblock Numerical integration of {E}instein's field equations.
\newblock {\em Phys. Rev. D (3)}, 59(2):024007, 7, 1999.

\bibitem{RRS98}
Gerhard Rein, Alan~D. Rendall, and Jack Schaeffer.
\newblock Critical collapse of collisionless matter - a numerical
  investigation.
\newblock {\em Phys. Rev. D}, 58:044007, 1998.

\bibitem{OlCh01}
I.~Olabarrieta and M.~Choptuik.
\newblock Critical phenomena at the threshold of black hole formation for
  collisionless matter in spherical symmetry.
\newblock gr-qc/0107076.

\bibitem{Rein93}
Gerhard Rein.
\newblock Static solutions of the spherically symmetric {V}lasov-{E}instein
  system.
\newblock {\em Math. Proc. Cambridge Philos. Soc.}, 115(3):559--570, 1994.

\bibitem{CP}
Frans Pretorius and Matthew~W. Choptuik.
\newblock Gravitational collapse in $2+1$ dimensional {A}d{S} spacetime.
\newblock {\em Phys. Rev. D (3)}, 62(12):124012, 15, 2000.

\bibitem{HO}
Viqar Husain and Michel Olivier.
\newblock Scalar field collpase in three-dimensional {A}d{S} spacetime.
\newblock {\em Classical Quantum Gravity}, 18(2):L1--L9, 2001.

\bibitem{gar}
David Garfinkle.
\newblock Exact solution for (2+1)-dimensional critical collapse.
\newblock {\em Phys. Rev. D (3)}, 63:044007, 5, 2001.

\bibitem{clement:quasi}
G.~Cl\'ement and A.~Fabbri.
\newblock Critical collapse in 2+1 dimensional ads spacetime: quasi-css
  solutions and linear perturbations.
\newblock gr-qc/0109002., 2001.

\bibitem{BTZ}
M{\'a}ximo Ba\~{n}ados, Claudio Teitelboim, and Jorge Zanelli.
\newblock Black hole in three-dimensional spacetime.
\newblock {\em Phys. Rev. Lett.}, 69(13):1849--1851, 1992.

\bibitem{Dain99}
Sergio Dain and Helmut Friedrich.
\newblock Asymptotically flat initial data with prescribed regularity.
\newblock {\em Comm. Math. Phys.}, 222(3):569--609, 2001.
\newblock gr-qc/0102047.

\bibitem{Beig80}
R.~Beig and W.~Simon.
\newblock Proof of a multipole conjecture due to {G}eroch.
\newblock {\em Comm. Math. Phys.}, 78(1):75--82, 1980/81.

\bibitem{Dain01b}
Sergio Dain.
\newblock Initial data for stationary space-time near space-like infinity.
\newblock {\em Class. Quantum Grav.}, 18(20):4329--4338, 2001.
\newblock gr-qc/0107018.

\bibitem{Winicour80}
Jeffrey Winicour.
\newblock Angular momentum in general relativity.
\newblock In {\em General relativity and gravitation, Vol. 2}, pages 71--96.
  Plenum, New York, 1980.

\bibitem{Moreschi98}
Osvaldo~M. Moreschi and Sergio Dain.
\newblock Rest frame system for asymptotically flat space-times.
\newblock {\em J. Math. Phys.}, 39(12):6631--6650, 1998.

\bibitem{Dain00}
Sergio Dain and Osvaldo~M. Moreschi.
\newblock General existence proof for rest frame systems in asymptotically flat
  spacetime.
\newblock {\em Classical Quantum Gravity}, 17(18):3663--3672, 2000.

\bibitem{date:killing}
G.~Date.
\newblock Isolated horizon, {K}illing horizon and event horizon.
\newblock {\em Classical Quantum Gravity}, 18(23):5219--5225, 2001.
\newblock gr-qc/0107039.

\bibitem{date:isolated}
G.~Date.
\newblock Notes on isolated horizons.
\newblock {\em Classical Quantum Gravity}, 17(24):5025--5045, 2000.

\bibitem{klein:prl2}
C.~Klein and O.~Richter.
\newblock Exact relativistic gravitational field of a stationary
  counterrotating dust disk.
\newblock {\em Phys. Rev. Lett.}, 83:2884--2887, 1999.

\bibitem{klein:prd4}
J.~Frauendiener and C.~Klein.
\newblock Exact relativistic treatment of stationary counterrotating dust
  disks: physical properties.
\newblock {\em Phys. Rev. D (3)}, 63(8):084025, 17, 2001.

\bibitem{neugebauermeinel1}
G.~Neugebauer and R.~Meinel.
\newblock General relativistic gravitational field of a rigidly rotating disk
  of dust: solution in terms of ultraelliptic functions.
\newblock {\em Phys. Rev. Lett.}, 75(17):3046--3047, 1995.

\bibitem{pfister1}
Urs~M. Schaudt and Herbert Pfister.
\newblock Isolated {N}ewtonian dust stars are unstable but can be stabilized by
  exterior matter.
\newblock {\em Gen. Relativity Gravitation}, 33(5):719--737, 2001.

\bibitem{dyson}
Freeman~J. Dyson.
\newblock Feynman's proof of the {M}axwell equations.
\newblock {\em Amer. J. Phys.}, 58(3):209--211, 1990.

\bibitem{singh:dadhich2}
Parampreet Singh and Naresh Dadhich.
\newblock The field equation from {N}ewton's law of motion and the absence of
  magnetic monopole.
\newblock {\em Int. J. Mod. Phys. A}, 16:1237--1247, 2001.

\bibitem{singh:dadhich}
Parampreet Singh and Naresh Dadhich.
\newblock Field theories from the relativistic law of motion.
\newblock {\em Modern Phys. Lett. A}, 16(2):83--90, 2001.

\bibitem{barbour:murch}
J.~Barbour, B.~Foster, and N.~\'O Murchadha.
\newblock Relativity without relativity.
\newblock gr-qc/0012089.

\bibitem{rendall:kichenassamy}
Satyanad Kichenassamy and Alan~D. Rendall.
\newblock Analytic description of singularities in {G}owdy spacetimes.
\newblock {\em Classical Quantum Gravity}, 15(5):1339--1355, 1998.

\bibitem{andersson:rendall:quiescent}
Lars Andersson and Alan~D. Rendall.
\newblock Quiescent cosmological singularities.
\newblock {\em Comm. Math. Phys.}, 218(3):479--511, 2001.

\bibitem{narita:etal:gowdy}
Makoto Narita, Takashi Torii, and Kengo Maeda.
\newblock Asymptotic singular behaviour of {G}owdy spacetimes in string theory.
\newblock {\em Classical Quantum Gravity}, 17(22):4597--4613, 2000.

\bibitem{damour:henneaux:homog}
T.~Damour and M.~Henneaux.
\newblock Oscillatory behaviour in homogeneous string cosmology models.
\newblock {\em Phys.Lett.}, B488:108--116, 2000.
\newblock hep-th/0006171.

\bibitem{nolan}
Brien~C. Nolan.
\newblock Sectors of spherical homothetic collapse.
\newblock {\em Classical Quantum Gravity}, 18(9):1651--1675, 2001.

\bibitem{beesham}
S.~G. Ghosh, R.~V. Saraykar, and A.~Beesham.
\newblock Collapsing shells of radiation in higher dimensional space-time and
  the cosmic censorship conjecture.
\newblock gr-qc/0106083, 2001.

\bibitem{konkowski}
T.~M. Helliwell and D.~A. Konkowski.
\newblock The strengths and limitations of a stability conjecture for {C}auchy
  horizons.
\newblock {\em Bull. Calcutta Math. Soc.}, 91(1):49--58, 1999.

\bibitem{andersson:edgar1}
Fredrik Andersson and S.~Brian Edgar.
\newblock Existence of {L}anczos potentials and superpotentials for the {W}eyl
  spinor/tensor.
\newblock {\em Classical Quantum Gravity}, 18(12):2297--2304, 2001.

\bibitem{andersson:edgar2}
F.~Andersson and S.~B. Edgar.
\newblock Local existence of symmetric spinor potentials for symmetric
  $(3,1)$-spinors in {E}instein space-times.
\newblock {\em J. Geom. Phys.}, 37(4):273--290, 2001.

\bibitem{senovilla}
Jos{\'e} M.~M. Senovilla.
\newblock Super-energy tensors.
\newblock {\em Classical Quantum Gravity}, 17(14):2799--2841, 2000.

\bibitem{gt}
Robert Geroch and Jennie Traschen.
\newblock Strings and other distributional sources in general relativity.
\newblock {\em Phys. Rev. D (3)}, 36(4):1017--1031, 1987.

\bibitem{cvw}
C.~J.~S. Clarke, J.~A. Vickers, and J.~P. Wilson.
\newblock Generalized functions and distributional curvature of cosmic strings.
\newblock {\em Classical Quantum Gravity}, 13(9):2485--2498, 1996.

\bibitem{penrose}
Michael Kunzinger and Roland Steinbauer.
\newblock A note on the {P}enrose junction conditions.
\newblock {\em Classical Quantum Gravity}, 16(4):1255--1264, 1999.

\bibitem{geo2}
M.~Kunzinger and R.~Steinbauer.
\newblock A rigorous solution concept for geodesic and geodesic deviation
  equations in impulsive gravitational waves.
\newblock {\em J. Math. Phys.}, 40(3):1479--1489, 1999.

\bibitem{vickersESI}
James~A. Vickers.
\newblock Nonlinear generalised functions in general relativity.
\newblock In {\em Nonlinear theory of generalized functions (Vienna, 1997)},
  pages 275--290. Chapman \& Hall/CRC, Boca Raton, FL, 1999.

\bibitem{found}
M.~Grosser, E.~Farkas, M.~Kunzinger, and R.~Steinbauer.
\newblock {\em On the foundations of nonlinear generalized functions {I},
  {II}.}, volume 153, number 729 of {\em Mem.\ Amer.\ Math.\ Soc.}
\newblock Amer.\ Math.\ Soc., 2001.

\bibitem{vim}
M.~Grosser, M.~Kunzinger, R.~Steinbauer, and J.~Vickers.
\newblock A global theory of algebras of generalized functions.
\newblock to appear.

\bibitem{ndg}
M.~Kunzinger and R.~Steinbauer.
\newblock Foundations of a nonlinear distributional geometry.
\newblock Acta Appl. Math.
\newblock to appear.

\bibitem{vw1}
J.~A. Vickers and J.~P. Wilson.
\newblock Generalized hyperbolicity in conical spacetimes.
\newblock {\em Classical Quantum Gravity}, 17(6):1333--1360, 2000.

\bibitem{vw2}
J.~A. Vickers and J.~P. Wilson.
\newblock Generalised hyperbolicity: hypersurface singularities.
\newblock gr-qc/0101018, 2001.

\bibitem{scott:szekeres}
Susan~M. Scott and Peter Szekeres.
\newblock The abstract boundary---a new approach to singularities of manifolds.
\newblock {\em J. Geom. Phys.}, 13(3):223--253, 1994.

\bibitem{mars}
Marc Mars.
\newblock Spacetime {E}hlers group: transformation law for the {W}eyl tensor.
\newblock {\em Classical Quantum Gravity}, 18(4):719--738, 2001.

\bibitem{yasuno}
Katsuhito Yasuno, Tatsuhiko Koike, and Masaru Siino.
\newblock Thurston's geometrization conjecture and cosmological models.
\newblock {\em Classical Quantum Gravity}, 18(8):1405--1420, 2001.

\end{thebibliography}

\end{document}